\documentclass[pre,floats,superscriptaddress,showpacs,usenames]{revtex4}
\usepackage{amsmath}
\usepackage{graphicx}

\usepackage{amsmath}
\usepackage{amsbsy}
\usepackage[normalem]{ulem}
\usepackage{amsfonts}

\begin{document}

\title{Consistent description of fluctuations requires negative temperatures.}

\author{Luca Cerino}
\email{luca.cerino@roma1.infn.it}
\affiliation{Dipartimento di Fisica, Universit\`a La Sapienza, and CNR - ISC, p.le A. Moro 2, 00185 Rome, Italy}

\author{Andrea Puglisi}
\email{andrea.puglisi@roma1.infn.it}
\affiliation{CNR-ISC and 
Dipartimento di Fisica, Universit\`a La Sapienza, p.le A. Moro 2, 00185 Rome, Italy}

\author{Angelo Vulpiani}
\email{angelo.vulpiani@roma1.infn.it}
\affiliation{Dipartimento di Fisica, Universit\`a La Sapienza, and CNR - ISC, p.le A. Moro 2, 00185 Rome, Italy}

\begin{abstract}
We review two definitions of temperature in statistical mechanics,
$T_B$ and $T_G$, corresponding to two possible definitions of entropy,
$S_B$ and $S_G$, known as surface and volume entropy respectively. We
restrict our attention to a class of systems with bounded energy and
such that the second derivative of $S_B$ with respect to energy is
always negative: the second request is quite natural and holds in systems of
obvious relevance, i.e. with a number $N$ of degrees of freedom
sufficiently large (examples are shown where $N \sim 100$ is sufficient)
and without long-range interactions. We first
discuss the basic role of $T_B$, even when negative, as the
parameter describing fluctuations of observables in a
sub-system. Then, we focus on how $T_B$ can be measured dynamically,
i.e. averaging over a single long experimental trajectory. On the
contrary, the same approach cannot be used in a
generic system for $T_G$, since the equipartition theorem may be
spoiled by boundary effects due to the limited energy. These general
results are substantiated by the numerical study of a Hamiltonian
model of interacting rotators with bounded kinetic energy. The
numerical results confirm that the kind of configurational order
realized in the regions at small $S_B$, or equivalently at small
$|T_B|$, depends on the sign of $T_B$.
\end{abstract}
\maketitle

\section{Introduction}
\label{Introduction}

Two different definitions of temperature in equilibrium statistical mechanics have
been recently the subject of an intense
debate~\cite{Dunkel2013,Vilar2014,Schneider2014,DunkelHil2014,Wang2014,Frenkel2014,Dunkel2014,Hilbert2014,Campisi2015,buonsante2015},
after the publication of experimental measurements of a negative
absolute temperature~\cite{Braun2013,Carr2013}.  In~\cite{Braun2013}
it was demonstrated the possibility to prepare a state where the
observed distribution of the modified kinetic energy {\em per atom} appeared to
be inverted, i.e. with the largest population in the high energy
states, yielding a {\em de facto} negative absolute temperature.

The possibility of a negative absolute temperature is well known since
the theoretical work by Onsager on the statistical hydrodynamics of
point vortices~\cite{Onsager1949} and the experimental and theoretical
results on nuclear spin systems by Pound, Ramsey and Purcell
(see~\cite{Ramsey1956,landsberg77,landsberg} for a review and discussion). In those  investigations, it
was clear that an inverse temperature parameter $\beta$
ranging in the full infinite real line $(-\infty,\infty)$ did not lead
to any  inconsistency or paradox. Ramsey in 1956 already
realised that ``the Carath\'eodory form of the second law is
unaltered.''~\cite{Ramsey1956}

A negative absolute temperature appears whenever the microcanonical
entropy is non-monotonic in the energy, a condition which can be
realized when the total energy has a global maximum, which may happen when the phase space is bounded. There are also
cases where the phase space is bounded but the energy diverges:
again this may lead to a non-monotonic entropy, an important example is given by point
vortices~\cite{Onsager1949,berdichevsky91,montgomery91,berdichevsky93,oneil93,berdichevsky95}. It is crucial to highlight that the lack of monotonicity (for entropy vs. energy) is realised if
one adopts the simplest definition of microcanonical entropy, which is
related to the logarithm of the number of states with a given energy. Since such a definition
appears in the so-called ``tombstone formula'' written on
Boltzmann's grave, ``$S=k \log W$'', it is often referred to
as Boltzmann's definition of entropy. Even if not historically
precise~\cite{Hilbert2014}, we adopt the same convention (but setting $k=1$) and call
``Boltzmann entropy'' of a system with Hamiltonian $H({\bf Q}, {\bf P})$ -- where ${\bf Q}$ and ${\bf P}$ are vectors in $\mathbb{R}^{dN}$, being $d$ the dimension of the system --
\begin{equation} \label{bent}
S_B(E,N) = \log \omega(E),
\end{equation}
being $\omega(E)$ the density of states, i.e. 
\begin{equation}
\omega(E) = \int \delta(H-E) d^{dN}Q d^{dN} P=\frac{\partial \Sigma(E)}{\partial E},
\end{equation}
and $\Sigma(E)$ the total ``number'' of states with energy less or equal
then $E$, that is
\begin{equation}
\Sigma(E) = \int_{H<E}d^{dN}Q d^{dN} P.
\end{equation}
In definition~\eqref{bent} we have ignored an additive constant which
is not relevant in our discussion. In~\cite{Hilbert2014} it is stated that the validity of the
second principle of thermodynamics depends on the value of this arbitrary
constant. Nonetheless, such an arbitrariness and the consequent paradox
can be removed if all the quantities (energies, positions, momenta, time etc...)
are considered adimensional.  Propagating the denomination, it is
customary to define the ``Boltzmann temperature'' through
\begin{equation} \label{btemp}
\beta_B=\frac{1}{T_B} = \frac{\partial S_B(E,N)}{\partial E}.
\end{equation}

Some authors~\cite{Dunkel2013,Hilbert2014} have argued that a
different definition of microcanonical entropy, proposed by Gibbs, has to be used in
statistical mechanics, in order to be consistent with a series of
``thermodynamic'' requirements and avoid unpleasant paradoxes. The Gibbs entropy, which is always monotonically increasing, reads
\begin{equation} \label{gent}
S_G(E,N) = \log \Sigma(E),
\end{equation}
and leads to the Gibbs temperature definition, which is always positive:
\begin{equation} \label{gtemp}
\beta_G=\frac{1}{T_G} = \frac{\partial S_G(E,N)}{\partial E} \ge 0.
\end{equation}

Let us note that, since $T_B$ is defined directly on the surface of
interest (i.e. that at constant energy $E$), from the point of view of
the ergodic approach its use appears rather natural. The Gibbs
temperature, on the other side, enters through an ensemble average in
the equipartition formula of textbooks~\cite{huang1988}:
\begin{equation} \label{eq:equip}
  \left\langle x_i \frac{\partial H}{\partial x_j} \right\rangle = \delta_{ij} T_G,
\end{equation}
where $x_i$ is any of the components of vector $({\bf Q},{\bf P})$ and
the average is done in the microcanonical ensemble. In
Section~\ref{dynamics}, we will discuss the limits of application of
formula~\eqref{eq:equip} when the energy is bounded. We also mention
that $T_G$ appears in the theory of Helmholtz monocycles (which had an
important role in the development of the Boltzmann's ideas for the
ergodic theory), for one-dimensional
systems~\cite{Helmholtz1884,Campisi2009}.

In spite of the fact that, in our opinion, the basic features of the
different definitions of temperature do not present particular
technical or conceptual subtleties, there is a certain confusion in
the literature; therefore a general discussion of the topic can be
useful. In this paper we present a line of reasoning where Boltzmann
temperature $T_B$ (positive or negative) is the (unique) proper
parameter which is relevant for the statistical properties of the
energy fluctuations, as well as in determining the flux of energy
between two  systems at different temperatures, in addition it is measurable, without
the appearance of any evident inconsistency. Let us remark that the
systems discussed in~\cite{Hilbert2014}, from which the authors try to
show that only $T_G$ is the ``good'' temperature, are small
($N=\mathcal{O}(1)$) and/or with long interactions.

In Section~\ref{concave}, after presenting the class of physically
relevant systems which are the subject of our study, we describe how
the Boltzmann temperature $T_B$ naturally describes fluctuations of
observables in subsystems, in analogy with the derivation of the
canonical ensemble from the microcanonical one. In
Section~\ref{dynamics} we discuss dynamical (``ergodic'')
measurements, which can reproduce $T_B$ but are in general unsuited to
measure $T_G$: in particular we show a possible failure of the
equipartition theorem.  In Section~\ref{model} we report a series of
numerical results with a model of interacting rotators with bounded
kinetic energy, discussing the many practical uses of Boltzmann
temperature. Summary and conclusions are drawn in
Section~\ref{conclusions}, together with a critique of some of the
arguments used, in~\cite{Hilbert2014}, to rule out the thermodynamic
meaning of $T_B$.

\section{The relevance of the Boltzmann temperature}
\label{concave}

In this section we show, following the standard approach that can be
found even in some textbooks, the unavoidable role of $T_B$ in many
problems of statistical mechanics.

\subsection{Systems of physical relevance}

In the rest of the paper we consider systems made of a finite but
large number $N \gg 1$ of particles with local interactions, i.e. we
exclude long-range potentials or mean-field models. It should be
understood that long-range interactions certainly widen the
phenomenology of statistical mechanics and may lead to complicate
functional dependences for $S_B(E,N)$, e.g. with several maxima or
minima, even for large $N$. Nevertheless they are not necessary for
the discussion of negative temperature and, most importantly, they
represent quite a peculiar case where even thermodynamics is not
obvious: for instance, it is not evident that the typical {\em
  Gedankenexperiment} of putting in contact two -- previously isolated
-- systems can be realized, as the isolation condition is
prevented by the long-range interaction.

We also assume that $S_B(E,N)$ is always convex, i.e. $d^2 S_B(E,N)
/dE^2 \le 0$. This is certainly true in the limit of vanishing
interaction and in short-range-interacting systems for large $N$,
since $S_B$ is strictly related to the large deviation function
associated to the density of states~\footnote{ It is interesting to
  notice that Kubo in~\cite{kubo65} uses the adjective ``normal'' for
  systems satisfying $\Sigma(E,N) \sim e^{N \phi(E/N) +o(N)}$.  It is
  easy to verify that for such systems one has
  $\beta_G=\beta_B+O(1/N)$.  However our assumption is different: we
  ask that, in the large $N$ limit, $\omega(E,N) \sim e^{N \psi(E/N)
    +o(N)}$.  Since $\Sigma(E,N)=\int^E \omega(E') dE'$, a simple
  steepest descend computation shows that, if $d \psi(E'/N)/dE'>0$ for
  $E'<E$, then $\psi(E/N)=\phi(E/N)$: this is equivalent to say that
  $T_B=T_G$ in the thermodynamic limit (i.e. up to $O(1/N)$) whenever
  $T_B>0$ (see Fig.~\ref{fig:phase_space} for an example).  On the
  other hand if $\psi$ has a maximum at $E^*$ then $\Sigma(E,N)$ is
  roughly constant for $E>E^*$. In summary, for ``normal'' systems the
  temperatures must coincide, while with our assumption, one can have
  different temperatures in the region $E>E^*$. Note also that normal
  systems also satisfy our assumption, while the opposite is not
  true. Moreover, even if not all the systems satisfying our
  assumption could be named ``normal'', all of them satisfy the
  equivalence of ensembles (as discussed below). }. Let us stress that
these large values of $N$ are not necessarily ``thermodynamic'' ($N
\to \infty$): for instance in Sec.~\ref{model} we will exhibit a
system that possesses all the required features already at $N=100$.
In general such a value of $N$ will depend on the specific system,
corresponding to situations in which some common approximations
(e.g. Laplace approximation for exponential integrals) can be safely
applied. In Sec. II.C we discuss in some details the origin of the
convexity of $S_B(E,N)$. It is easy to understand that this assumption
implies the validity of the second principle of thermodynamics, as
discussed in the next subsection.

\subsection{Second law and energy flux between two systems in contact}
Let us consider a system $\mathcal{A}$ of $N_\mathcal{A}$ particles
described by the variables
$\{ {\bf Q}_\mathcal{A},{\bf P}_\mathcal{A}\}$ and Hamiltonian
$H_\mathcal{A}({\bf Q}_\mathcal{A},{\bf P}_\mathcal{A})$, a
system $\mathcal{B}$ of $N_\mathcal{B}$ particles described by the
variables $\{ {\bf Q}_\mathcal{B},{\bf P}_\mathcal{B}\}$ and
Hamiltonian
$H_\mathcal{B}({\bf Q}_\mathcal{B},{\bf P}_\mathcal{B})$ and
a small coupling among the two, so that the global Hamiltonian is
\begin{equation}
  H=H_\mathcal{A}( {\bf Q}_\mathcal{A},{\bf P}_\mathcal{A})+H_\mathcal{B}({\bf Q}_\mathcal{B},{\bf P}_\mathcal{B})+H_I({\bf Q}_\mathcal{A}, {\bf Q_\mathcal{B}}).
\end{equation}
If the two Hamiltonians have the same functional dependencies on the
canonical variables (i.e. they correspond to systems with same
microscopic dynamics, with possibly different sizes $N_A$
and $N_B$), for large $N$, we
can introduce the (Boltzmann) entropy per particle
\begin{equation}
S_B(E,N)=N S(e) \,\,\,\, , \,\,\,\, e=\frac{E}{N},
\end{equation}
with $S(e)$ a convex function, identical for systems $\mathcal{A}$ and
$\mathcal{B}$. Let us now suppose that systems $\mathcal{A}$ and
$\mathcal{B}$ have, respectively, energy
$E_\mathcal{A}=N_\mathcal{A} e_\mathcal{A}$ and
$E_\mathcal{B}=N_\mathcal{B} e_\mathcal{B}$ and the corresponding
inverse Boltzmann temperatures $\beta_B^{(\mathcal{A})}$ and
$\beta_B^{(\mathcal{B})}$.\\
 When the two systems are put in contact, a
new system is realized with $N=N_\mathcal{A}+N_\mathcal{B}$
particles. Let us call $a=N_\mathcal{A}/N$ the fraction of particles
from the system $\mathcal{A}$. We have that the final energy is
$E_f=E_\mathcal{A}+E_\mathcal{B}=N e_f$, where
$e_f=a e_\mathcal{A}+(1-a)e_\mathcal{B}$ and final entropy
\begin{equation}
  S_B(E_f,N)=N S(e_f)\ge N_\mathcal{A}S(e_1)+N_\mathcal{B} S(e_\mathcal{B})= N [a S(e_\mathcal{A}) +(1-a) S(e_\mathcal{B})].
\end{equation}
The previous inequality follows from the convexity assumption for
$S(e)$ which implies
\begin{equation}
S(ae_\mathcal{A}+ (1-a)e_\mathcal{B})\ge a S(e_\mathcal{A}) +(1-a) S(e_\mathcal{B}).
\end{equation}
The final inverse temperature $\beta_B^{(f)}$ is intermediate between
$\beta_B^{(\mathcal{A})}$ and $\beta_B^{(\mathcal{B})}$, e.g. if
$e_\mathcal{B} >e_\mathcal{A}$ -- that is
$\beta_B^{(\mathcal{A})}>\beta_B^{(\mathcal{B})}$ -- then
\begin{equation}
\beta_B^{(\mathcal{B})}<\beta^{(f)}<\beta_B^{(\mathcal{A})}.
\end{equation}
The energy flux obviously goes from smaller $\beta_B$ (hotter) to
larger $\beta_B$ (colder). The consequence of convexity is that
$\beta_B(E)$ is always decreasing and a negative value does not lead
to any ambiguity. Confusion may arise from the fact that $T_B<0$ is,
for the purpose of establishing the energy flux, hotter than
$T_B>0$. However if $\beta_B$ is used, the confusion is totally
removed~\cite{Ramsey1956}.\\
We also briefly discuss a particularly interesting case with different Hamiltonians. Suppose that for
the system $\mathcal{A}$ negative temperatures can be present, whereas
system $\mathcal{B}$ has only positive temperatures; it is quite easy
to see that the coupling of the system $\mathcal{A}$ at negative
temperature with the system $\mathcal{B}$ at positive temperature
always produces a system with final positive temperature.  Indeed, at
the initial time the total entropy is
\begin{equation}
S_{I}=S^\mathcal{A}(E_\mathcal{A})+S^\mathcal{B}(E_\mathcal{B}),
\end{equation}
while, after the coupling, it will be 
\begin{equation}
S_{F}=S^\mathcal{A}(E'_\mathcal{A})+S^\mathcal{B}(E'_\mathcal{B}),
\end{equation}
where $E'_\mathcal{A}+E'_\mathcal{B}=E_\mathcal{A}+E_\mathcal{B}$ and, within our assumptions, $E'_\mathcal{A}$ is
determined by the equilibrium condition~\cite{huang1988} that $S_{F}$ takes the maximum possible
value, i.e.
\begin{equation}
\beta_\mathcal{A}={\partial S^\mathcal{A}(E'_\mathcal{A}) \over \partial E'_\mathcal{A}}=
\beta_\mathcal{B}={\partial S^\mathcal{B}(E'_\mathcal{B}) \over \partial E'_\mathcal{B}}.
\end{equation}
Since $\beta_\mathcal{B}$ is positive for every value of
$E_\mathcal{B}'$, the final common temperature must also be
positive. The above conclusion can also be found, without a detailed
reasoning, in some textbooks
\cite{toda1992statistical,callen2006thermodynamics}.
\subsection{Subsystems}

\label{sec:sub}

Let us consider a vector ${\bf X}$ in $\mathbb{R}^{2d N_1}$ (with $N_1<N$), that is a subsystem of the full phase space
$({\bf Q}, {\bf P})$, and let us indicate with  $\widetilde {\bf X}$ in $\mathbb{R}^{2d(N-N_1)}$
the remaining variables. We have
\begin{equation}
H=H_1({\bf X})+H_2(\widetilde{\bf X})+H_I({\bf X}, \widetilde{\bf X})
\end{equation}
with an obvious meaning of symbols.

Let us consider the case $N\gg 1$ and $N_1\ll N$. In the microcanonical ensemble with energy $E$, the probability density function (pdf) for the full phase space
$({\bf Q}, {\bf P})$ is
\begin{equation}
P({\bf Q}, {\bf P})=\frac{1}{\omega(E,N)} \delta(H({\bf Q}, {\bf P})-E).
\end{equation}
The pdf of ${\bf X}$ can be obtained from the latter, by integrating over $\widetilde{\bf X}$. If the Hamiltonian
$H_I({\bf X}, \widetilde{\bf X})$ is negligible (a consequence of our assumption for non long-range interaction) then we have
\begin{equation} \label{subpdf}
P({\bf X})\simeq \frac{\omega(E-H_1({\bf X}), N-N_1)}{\omega(E,N)}. 
\end{equation}
It is now possible to exploit the definition of $S_B$ and get
\begin{align}
\omega(E,N) &= e^{S_B(E,N)}\\
\omega(E-H_1({\bf X}),N-N_1) &= e^{S_B(E-H_1({\bf X}),N-N_1)} \propto e^{S_B(E,N-N_1)-\beta_B(E)H_1({\bf X})},
\end{align}
which, together with~\eqref{subpdf} leads to
\begin{equation}\label{eq:single_particle}
P({\bf X})\propto  e^{-\beta_B H_1({\bf X})}.
\end{equation}
When $H_1$ is bounded (as in our assumptions), the previous
simple derivation can be done irrespective of the sign of $\beta_B$.
It is immediately clear from the above argument that $T_B$ is the
temperature ruling the statistics of fluctuations of physical
observables in a subsystem.  For instance, the pdf of the subsystem (i.e. the canonical ensemble)
energy $E_1$ reads
\begin{equation} \label{suben}
P(E_1,N_1) \propto \, \omega(E_1,N_1) e^{-\beta_B E_1} \propto e^{[S_B(E_1,N_1)-\beta_B E_1]}.
\end{equation}
Of course the above result holds in the (important) case where the
two subsystems are weakly interacting and $H_1 \ll E$.
Therefore, for $e_1=E_1/N_1$, one has
\begin{equation} \label{suben2}
P(e_1,N_1) \propto e^{N_1[S(e_1)-\beta_B e_1]},
\end{equation}
which is a large deviation law where the Cramer's function $C(e_1)$ is
$C(e_1)=\beta_B e_1-S(e_1)+\textrm{const}$. From general arguments of theory of probability, we know that - if
a large deviation principle holds - $\frac{d^2 C(e_1)}{de_1^2}\ge 0$
so $\frac{d^2S(e_1)}{d e_1^2} \le 0$. The validity of the large
deviation principle can be easily shown for non-interacting systems. For weakly
interacting systems it is quite common and reasonable, and can be stated under rigorous hypothesis~\cite{touchette2009,ld2014}.

\subsection{The generalised Maxwell-Boltzmann distribution}

\label{sec:mb}

The extreme case of the above considerations is when $N_1=1$, that is
to say the fluctuations of a single degree of freedom (e.g. a momentum
component of a single particle) are observed. This becomes interesting
when the Hamiltonian has the form
\begin{equation}
H=\sum_{n=1}^N g(p_n) + \sum_{n,k}^N V(q_n,q_k)
\end{equation}
where the variables $\{ p_n \}$ are limited and the same happens for the function $g(p)$.

Repeating the arguments in the previous subsection, one may compute the probability density for the distribution of a single momentum $p$, obtaining
\begin{equation} \label{maxbol}
P(p)\simeq \frac{\omega(E-g(p), N-1)}{\omega(E,N)} \propto  e^{-\beta_B g(p)},
\end{equation}
which, again, is valid for both positive and negative $\beta_B$.  We
mention that in the experiment in~\cite{Braun2013}, the
above recipe has been applied to measure both positive and negative system's
temperatures.

From Eqs.~\eqref{suben} and~\eqref{maxbol} the true deep meaning of the
(Boltzmann) temperature is quite transparent: it is a quantity which
rules the pdf of energy of a subsystem (or the momentum of a single
particle). Let us note that since $T_B$ is associated to the large
microcanonical system (in physical terms the reservoir) it is a
non-fluctuating quantity~\cite{Tfluct} also for each sub-system and, in general, for non-isolated systems. In the conclusions, we discuss again such an
aspect which is not always fully understood, see
e.g. Ref.~\cite{Hilbert2014}

\subsection{Temperature and order}

\label{sec:order}
In usual statistical mechanics, low temperatures -- or, better, high
values of inverse temperature -- are usually associated to the possibility of some kind of order, the
most noticeable example given by phase transitions. Intuitively, one
would expect such a situation whenever $\omega(E)$ is relatively
small, which usually corresponds to regions where $|\beta_B|$ is large
irrespective of the temperature's sign. A famous example where such an
order at negative (small) temperatures was observed is that of
pointlike vortices discussed by Onsager in~\cite{Onsager1949}. The system, obtained as a particular limit from
two-dimensional Euler equations, describes $N$ points of
vorticities $\{ \Gamma_1, ... ,\Gamma_N \}$ in a two-dimensional
domain $\Omega$: the equation of motions of the coordinates $(x_n,y_n)$ of the $n$-th point vortex are
shown to be (see for instance~\cite{cencini2010chaos})
\begin{equation}
\Gamma_i \frac{dx_i}{dt}=\frac{\partial H}{\partial y_i} \,\,\, , \,\,\,
\Gamma_i \frac{dy_i}{dt}=-\frac{\partial H}{\partial x_i}
\end{equation}
with Hamiltonian
\begin{equation}
H=\sum_{i\neq j} \Gamma_i \Gamma_j {\cal G}(r_{i,j})
\end{equation}
where ${\cal G}(r)$ is the Green function of the Laplacian in  $\Omega$: in the infinite plane one has
${\cal G}(r)= -1/4 \pi \ln r$ where $r_{i,j}=\sqrt{(x_i-x_j)^2+(y_i-y_j)^2}$. The canonical variables in this case are
\begin{equation}
q_i=   \sqrt{|\Gamma_i|}x_i
\,\,\, ,\,\,
p_i= \sqrt{|\Gamma_i|}\, \text{sign}(\Gamma_i)\,y_i
\end{equation}
Onsager showed that if the domain of $\Omega$ is bounded, then
negative $T_B$ are achieved at large values of the energy.  At large
energies a particular spatial order appears too: clusters of vortices
with the same sign of the vorticity are the structures most easily
found.  It is interesting to notice that $T_B<0$ (and the
corresponding clusterization) is not a peculiarity of the divergence
of $\mathcal{G}(r)$ in $r=0$, nor of the long range nature of the
interaction: indeed, it can be obtained with any arbitrary
${\cal G}(r)$ having a maximum (even finite) in $r=0$, and vanishing
at large $r$, provided that the domain is bounded. The presence of
spatial order at high values of energy, in the form of discrete
breathers, has been observed also in the discrete non-linear
Schr\"odinger equation and analogous systems
\cite{iubini2013discrete,buonsante2015}. In Section~\ref{model} we
introduce a different, in a way simpler, model which still exhibits
spatial order at small negative temperatures.

\section{How to measure $T_B$ and $T_G$}
\label{dynamics}


The definitions of $\beta_B$ and $\beta_G$ given in Eqs.~\eqref{btemp}
and~\eqref{gtemp} are based on the functional dependence of the phase
space occupations $\omega(E)$ and $\Sigma(E)$ upon the energy. In a
real or numerical experiment it may be cumbersome or even impossible
to make use of those definitions to measure the two temperatures: for
instance, an empirical estimate of $\omega(E)$ (and therefore of
$\Sigma(E)$) will always be limited by the available statistics
(number of independent measurements of $E$) and therefore cannot provide
a clear answer, for both $\beta_B$ and $\beta_G$, in the interesting regimes where $\omega(E) \sim 0$.

On the other hand it has been shown~\cite{rugh1997} that $\beta_B$ can be obtained as a microcanonical
average of a certain observable. The recipe is the following
\begin{equation} \label{rugh}
 \beta_B=<R({\bf X})> \,\,\, , \,\,\,\, R({\bf X})=\nabla \cdot \frac{\nabla H}{|\nabla H|^2}
\end{equation}
where $\nabla$ stands for the vector of derivative operators along the
degrees of freedom in the full phase space ${\bf X} \equiv ({\bf Q},
{\bf P})$. From~\eqref{rugh} one has, assuming the ergodicity, that
$\beta_B$ can be computed with a molecular dynamics simulation, and,
at least in principle, by a long-time series from an experiment.
It is interesting to notice that such a kind of recipe does
not exist for $S_B(E,N)$ or $S_G(E,N)$~\cite{rugh1997}. It is clear that, in view of the
considerations in Sections~\ref{sec:sub} and~\ref{sec:mb},
one may always measure fluctuations of appropriate observables, such
as subsystem's energy or single particle momentum, to get an estimate of $T_B$. 

Coming to $\beta_G$, a way, even discussed in textbooks and
considered sometimes rather important~\cite{Hilbert2014}, to approach
the problem of its measurement is via the equipartition theorem,
which states
\begin{equation} \label{eqpart}
\left\langle x_i \frac{\partial H}{\partial x_j} \right\rangle = \delta_{ij} T_G.
\end{equation}
However the usual derivation of Eq.~\eqref{eqpart} implies the
possibility to neglect boundary terms in an integration by parts. Such
a possibility is challenged in the class of systems with bounded energy and phase space 
that we are considering.

In particular it is easy to show that~\eqref{eqpart} does not hold under the simultaneous realization of the following conditions:
\begin{itemize}

\item bounded space of the canonical variables;

\item bounded derivatives of the Hamiltonian $ \frac{\partial H}{\partial x_j} $;

\item bounded energy from above and below: $E_m\le E \le E_M$;

\item vanishing density of states at the boundaries, i.e. $\omega(E_M)=0$.

\end{itemize}
Given such conditions, one has that, on one side,
\begin{equation}
T_G(E)=\frac{\Sigma(E)}{\omega(E)}
\end{equation}
diverges when $E \to E_M$. On the other side,
$\langle x_i \frac{\partial H}{\partial x_j} \rangle$ is limited,
resulting in a contradiction.

A failure or the equipartition formula Eq. \eqref{eqpart} is also possible in systems where there are no negative temperatures, i.e.
 $T_G \simeq T_B >0$ for all $E$. Consider, for instance, the following
Hamiltonian
\begin{equation}
H=\sum_{n=1}^N {p_n^2 \over 2} +\epsilon \sum_{n=1}^N ( 1- \cos (\phi_n-\phi_{n-1}) )
\end{equation}
where $\phi_n \in [-\pi , \pi)$.  For large $E$, i.e.
$E \gg \epsilon N$, the contribution to $\Sigma(E)$ of the variables
$\{ \phi_n \}$ does not depend too much on the value of $E$, so that
\begin{equation}
\Sigma_{\epsilon}(E)\simeq \Sigma_{0}(E)\propto \, E^{N/2} \,\, ,
\end{equation}
and  $T_G\simeq 2E/N$ and, for large $N$,  $T_B=T_G+O(1/N)$.

On the other hand it is easy to see that
\begin{equation}
\left|  \phi_n {\partial H \over \partial \phi_n} \right| \le 2 \pi \epsilon \,\, ,
\end{equation}
and, therefore, the equipartition formula
$\langle \phi_n {\partial H \over \partial \phi_n} \rangle =T_G$ does not
hold for large value of $E$ and $N$.


\section{Numerical results for a system with negative temperature}
\label{model}

In this Section we present a detailed study of a system composed of
$N$ ``rotators'' with canonical variables
$\phi_1,...,\phi_N,p_1...p_N$ with all $\phi_i$ and $p_i$ defined in
$[-\pi,\pi)$, and with Hamiltonian
\begin{equation} \label{modham}
  H(\phi_1,\ldots,\phi_N,p_1,\ldots,p_N)=\sum_{n=1}^N [1-
  \cos(p_n)] + \epsilon\sum_{n=1}^N [1 -\cos (\phi_n -\phi_{n-1})].
\end{equation}
Choosing, as boundary condition, $\phi_0=0$ guarantees that the only
conserved quantity by the dynamics is the total energy $E$. The
equations of motion for the rotators can be readily obtained applying
Hamilton's equations to Eq. (\ref{modham}):
\begin{eqnarray}\label{eq:motion}
\dot{\phi}_n&=& \sin(p_n),\nonumber\\
\dot{p}_n&=&-\epsilon\left( \sin(\phi_n-\phi_{n-1}) +\sin(\phi_n - \phi_{n+1})  \right).
\end{eqnarray}
\begin{figure}
\includegraphics[width=0.5\textwidth]{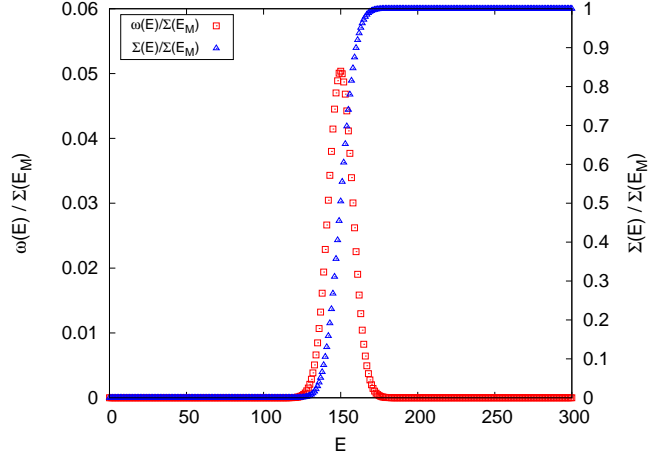}
\caption{Phase space sampling: we report the reconstruction of the
  density of states $\omega(E)$ and its integral $\Sigma(E)=\int_0^E
  dE' \omega(E')$. The two functions are normalized with
  $\Sigma(E_M=2N(1+\epsilon))$. The parameters of the system are:
  $N=100$ and $\epsilon=0.5$.}\label{fig:phase_space}
\end{figure} 
It is immediate to verify that the energy has a maximum value $E_M=2N(1+\epsilon)$ which is realised when $p_n=\pi$ and $\phi_n-\phi_{n-1}=\pi$ for every $n$.

When $\epsilon=0$ it is immediate to see that Hamiltonian in
Eq.~\eqref{modham} implies negative Boltzmann temperatures. Indeed at small
energy one has $1-\cos(p_n)\simeq p_n^2/2$ so that
\begin{equation}
\Sigma(E) \simeq C_N E^{N/2} \,\,\, , \,\, \omega(E) \simeq  \frac{N}{2} C_N E^{N/2-1}
\end{equation}
with $C_N=(2\pi)^N\frac{\pi^{N/2}}{\Gamma(N/2+1)}$. Close to $E_M=2N$ one has
$1-\cos(p_n)\simeq (\pi- p_n)^2/2$, therefore when $E$ approaches $E_M$ it is 
\begin{multline}
\Sigma(E)=\Sigma(E_M)-(2\pi)^N\int_{E<H<E_M} \prod_{n=1}^N dp_n
\simeq \Sigma(E_M)-(2\pi)^N\int_{\sum_n \frac{(\pi-p_n)^2}{2} < (E_M-E)} \prod_{n=1}^N dp_n=
\Sigma(E_M)-C_N(E_M-E)^{N/2}
\end{multline}
and therefore
\begin{equation}
\omega(E) \simeq  \frac{N}{2} C_N (E_M-E)^{N/2-1} \,\, .
\end{equation}
In conclusion we have that $\omega(E)=0$ if $E=0$ and $E=E_M$, which
implies a maximum in between and a region (at high energies) with
negative $\beta_B$.  The previous scenario is expected to hold also in the
prescence of a small interaction among the rotators and can be
numerically confirmed with a sampling of the phase-space (see
Fig. \ref{fig:phase_space}): random configurations of the system are
extracted with an uniform distribution over the phase space and
$\omega(E)$ is reconstructed by counting the number of configurations
lying in a small interval of width $\delta E$ around the energy $E$.
It is rather evident from Fig. \ref{fig:phase_space} that: the density of
states $\omega(E)$ has a maximum in $\tilde{E}\approx E_M /2$; it is
an increasing function for $E<\tilde{E}$ whence $T_B>0$; it decreases
for $E>\tilde{E}$ whence $T_B<0$.  Unfortunately, such a sampling is
reliable only in a narrow region around $\tilde{E}$: indeed, there are
very few configurations with energies much larger or smaller than
$\tilde{E}$ and, therfore, there is an extremely small probability to
extract such configurations with this procedure.

For this reason,  we have performed dynamical measures through numerical
simulations of the motion of the system: the integration of
Eqs. \eqref{eq:motion} is done with the usual Verlet scheme with a time step
$\Delta t=10^{-3}$.

\subsection{Measure of $T_B$}
\begin{figure}[!hbtp]
\includegraphics[width=0.5\textwidth]{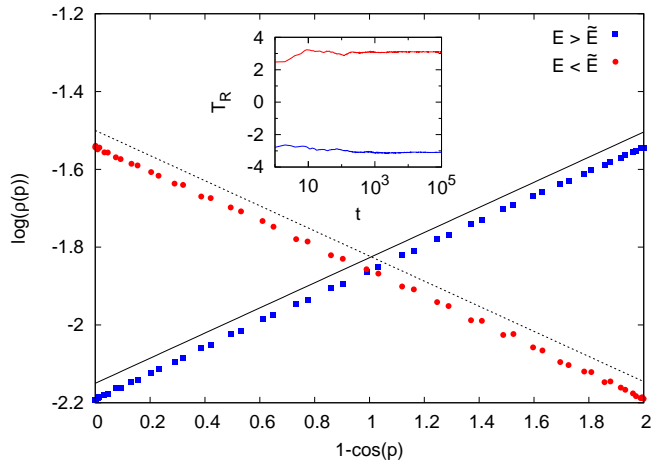}
\caption{Measure of the Boltzmann temperature in the rotators chain
  for $N=100$ and $\epsilon=0.5$. Probability distribution function of
  the momentum of the rotators as a function of their ``kinetic energy''
  $g(p)=1-\cos(p)$ at energy $E=E_-=170$ (blu squares) and
  $E=E_+=130$. The slopes of the two black straight lines are
  $1/T_R^\infty(E)$, where $T_R^\infty(E)$ is the asymptotic value of
  the corresponding curve in the inset. Inset: The $T_R$ obtained from
  the cumulated average of the observable $R({\bf X}(t))$ over a
  trajectory up to time $t$ at $E=170$ (blue line) and $E=130$ (red
  line). }\label{fig:boltz_temp}
\end{figure}

Measurements of the Boltzmann temperature are done with the two
methods discussed in the previous Sections. In particular, by
computing the following average (over a single trajectory of the
system)
\begin{equation}
\rho(p)=\lim_{\tau \to \infty} \frac{1}{N\tau} \int_0^\tau dt\,\, \sum_{i=1}^N \delta \left( p_i(t) -p\right),
\end{equation}
for different values of $p$, and assuming that the system is ergodic,
we recover the single-particle-momentum probability density function
$P(p)$, Eq. \eqref{eq:single_particle}. The result of such a measure
is reported in Fig. \ref{fig:boltz_temp}: for two different values of
energy $E_+<\tilde{E}$ and $E_->\tilde{E}$ the measured $\rho(p)$ is
plotted as a function of the ``kinetic energy'' of the individual rotator
$g(p)=1-\cos(p)$. The presence of a negative temperature at $E=E_-$
can be readily indentified by means of the consideration in Section
\ref{sec:mb}. Indeed, on one hand, the exponential behaviour of
$\rho(p)$ guarantees that the approximation used to obtain
Eq. \eqref{maxbol} is already valid (for every value of $g(p)$)
at $N=100$. On the other hand, the clear positive slope of the
function at $E=E_-$ is a direct consequence of the fact that
$T_B(E_-)<0$: the opposite situation is encountered at $E=E_+$, where
the decreasing behavior of $\rho(p)$ indicates a temperature
$T_B(E_+)>0$.  These conclusions can also be drawn by measuring the
time average of the function $R(X)$, Eq. \eqref{rugh}: in the inset of
Fig. \ref{fig:boltz_temp} we report the temperature obtained with the
cumulated average of $R({\bf X})$ up to time $t$, namely
\begin{equation}
\frac{1}{T_R(t)}=\frac{1}{t}\int_0^t\,\, dt'\,R({\bf X}(t')),
\end{equation}
for $E=E_+$ and $E=E_-$. These two quantities converge, for large $t$,
to an asymptotic value representing an estimate of the inverse
Boltzmann temperature $\beta_B$ of the system. This value, as
expected, is positive for $E=E_+$ and negative for $E=E_-$: moreover,
the values are in very good agreement with the slopes of the single particle
distribution function, as shown by the dashed and solid lines in
Fig. \ref{fig:boltz_temp}.

\subsection{Equivalence of ensembles and the equipartition formula}

Let us briefly discuss the problem of the equivalence of ensembles. In
the usual treatment of textbooks one starts from Eq. (23): assuming
that $S(e)$ is convex and performing a steepest descent analysis, for
large $N$, one obtains the canonical functions from the (Boltzmann) microcanonical ones, e.g.:
\begin{equation}
T_B(e) S(e) = e-f(T_B(e)),
\end{equation}
where $f(T)$ is the free energy per particle in the canonical ensemble. In addition
 the energy fluctuations are negligible. In such a derivation,
the relevant point is only the convexity of $S(e)$ and nothing about
its first derivative is asked. Therefore, the equivalence of ensembles
naturally holds under our hypothesis even for negative $T_B$. Since
$T_B$ and $T_G$ can be different even for large $N$, as in our model
defined with Eq. (35), it is evident that $T_G$ is not relevant for
the ensemble equivalence.

\begin{figure}
\includegraphics[width=0.5\textwidth]{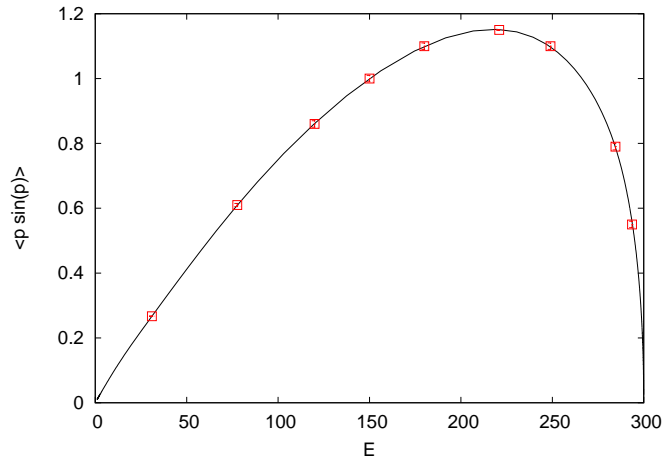}
\caption{Black line: $\langle p\sin(p)\rangle_\beta$ vs $U(\beta)$ in
  the canonical ensemble (Eqs. \eqref{eq:av_energy} and
  \eqref{eq:av_gibbs}) as parametric functions of
  $\beta\in(-\infty,\infty)$. Red squares: time averages of the
  equipartition function in molecular dynamics simulations at fixed
  energy $E$ (microcanonical ensemble). The values for the parameters
  of the model are $N=100$ and $\epsilon=0.5$.}\label{fig:gibbs}
\end{figure}

A common way \cite{Hilbert2014} to measure the Gibbs
temperature is by means of the equipartition formula,
Eq. \eqref{eqpart}: for the Hamiltonian in Eq. \eqref{modham} one should
get
\begin{equation} \label{micro}
\langle p_k \sin p_k \rangle_E =T_G(E),
\end{equation}
for every $1\le k \le N$. In the present subsection, we use the
notation $\langle \rangle_E$ to denote the average in the
microcanonical ensemble, in order to distinguish it from a canonical
average $\langle \rangle_\beta$ which is useful to get some analytic expressions and better investigate the validity of Eq.~\eqref{micro}.
The canonical probability density reads
\begin{equation}
\rho(\phi_1,\ldots,\phi_N,p_1,\ldots,p_n)=\frac{1}{Z(\beta)}e^{-\beta H(\phi_1,\ldots,\phi_N,p_1,\ldots,p_n)},
\end{equation}
where $Z(\beta)$ is the partition funcion and $\beta$ the (external)
inverse temperature, that can be either positive or negative: if such
a distribution is derived from a larger isolated system, as already
discussed in Section \ref{sec:sub}, the temperature in the canonical
ensemble is precisely the Boltzmann temperature of the whole system.
A simple explicit expression (see details of analogous calculations in
Ref. \cite{Livi1987}) can be derived for the mean energy
\begin{equation}\label{eq:av_energy}
  U(\beta)=\langle H\rangle_\beta=N \left(1+\epsilon  - \frac{I_1(\beta)}{I_0(\beta)}-\frac{\epsilon I_1(\beta \epsilon)}{I_0(\beta \epsilon)}\right),
\end{equation}
where $I_0(x)$ and $I_1(x)$ are, respectively, the zeroth and the
first modified Bessel function of the first kind. Analogously, one can
get an analytic formula for the equipartition function
\begin{equation}\label{eq:av_gibbs}
\langle p \sin(p)\rangle_\beta = \frac{1}{\beta}-\frac{e^{-\beta}}{\beta I_0(\beta)}.
\end{equation}
Let us remark that Eqs. \eqref{eq:av_energy} and \eqref{eq:av_gibbs}
hold for both positive and negative $\beta$. In
Fig. \ref{fig:gibbs} we report the plot of the parametric curve
$(U(\beta),\langle p \sin(p)\rangle_\beta)$ obtained by varying $\beta$ both in the
positive and in the negative
region of the real axis.

This curve is then compared with measures of $\langle p \sin(p)\rangle_E$ computed from
molecular dynamics simulations in the microcanonical ensemble at different
values of the energy $E$ (Fig. \ref{fig:gibbs}). Such a comparison
clearly shows that the results obtained in the two different ensembles
are identical, a transparent evidence that the equivalence of ensemble
already exists for this system quite far from the thermodynamic limit
($N=100$).

Fig. \ref{fig:gibbs} also shows that the equipartition
formula cannot be used to measure the Gibbs temperature: indeed, as
already pointed out in Section \ref{dynamics}, the equipartition
theorem can fail if the density of states $\omega(E)$ vanishes. This
is the case of our system (Fig. \ref{fig:phase_space}), where
$T_G=\Sigma(E)/\omega(E)$ should diverge for $E\to2 N(1+\epsilon)$: on
the other hand the results obtained in the canonical and in the
microcanonical ensemble clearly indicate that $\langle p \sin(p)\rangle_E\to 0$ as
$E\to 2N(1+\epsilon)$.

\subsection{Spatial coherence}

\begin{figure}[!hbtp]
\includegraphics[width=0.48\textwidth]{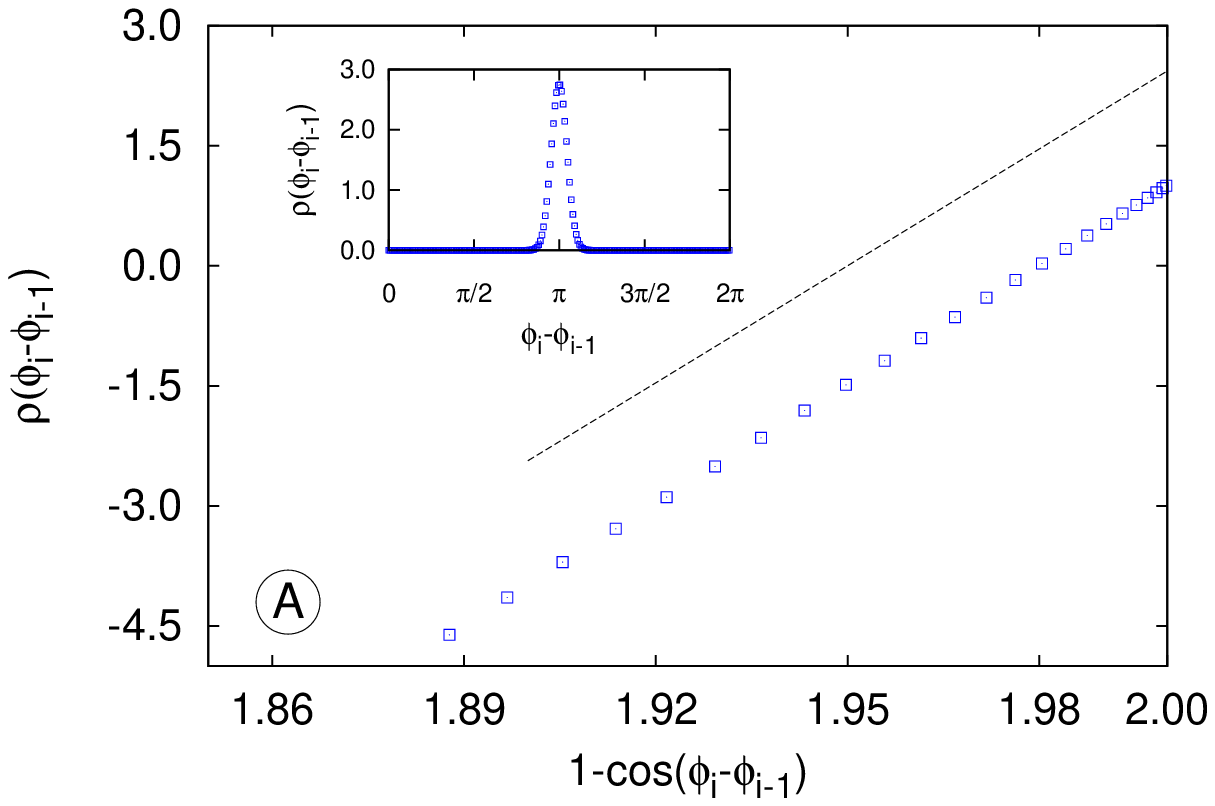}
\includegraphics[width=0.48\textwidth]{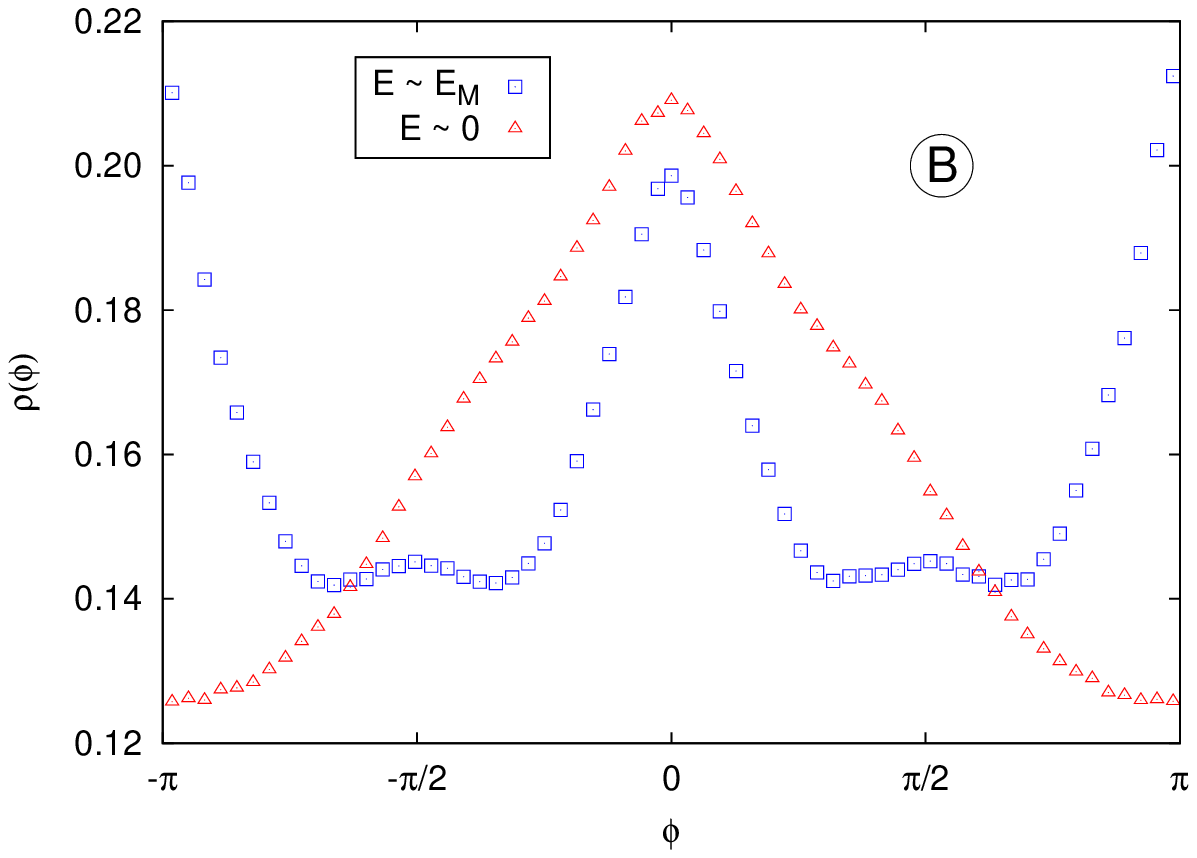}
\caption{A: Probability distribution function of angular distance
  between two consecutive rotators at high energy $E=298.96$. B:
  Probability distribution function of rotators' positions $\phi$ in
  the high energy case $E=298.96$ (blue squares) and in the low energy
  case $E=6.79$ (red triangles). The two maxima of the high energy
  distribution correspond to the clusters around $\phi=0$ and
  $\phi=\pi$ discussed in the text. The other parameters are $N=100$
  and $\epsilon=0.5$.}\label{fig:spatial_coherence}
\end{figure}

In analogy with systems of point vortices discussed in
Sec. \ref{sec:order}, the rotators model in Eq.\eqref{modham}
possesses a spatially ordered phase at large values of $E$: this can
be easily understood by noting that the density of states $\omega(E)$
vanishes in $E=E_M$, i.e. that there is a small number of microscopic
configurations corresponding to large values of $E$. In particular,
the maximum of the energy $E_M=2 N(1+\epsilon)$ is attained by the
unique microscopic state where, for every $n$, $p_n=\pi$ and
$\phi_n-\phi_{n-1}=\pi$; that is, where all the rotators are fixed
($\dot{\phi}=\sin \pi =0$) and the distance among two consecutive
rotators is $\Delta \phi=\pi$. As a consequence, since $\phi_0=0$, all
the particles with even index ($n=0,2,4\ldots$) must be at $\phi=0$
and the others ($n=1,3,\ldots)$ in $\phi=\pi$. At smaller values of
$E\lesssim E_M$, see Fig. \ref{fig:spatial_coherence} B, such
considerations can be extended, yielding a very similar situation:
even and odd rotators must be close, respectively, to $\phi=0$ or
$\phi=\pi$.

 Let us note that an ordered phase exists whenever, at a
given energy $E$, the number of corresponding configurations is small,
i.e. when $\omega(E)$ vanishes: for instance, the clustering can also
be observed at small energies, when the rotators accumulate around
$\phi=0$, in order to minimize the interaction energy, see
Fig. \ref{fig:spatial_coherence} B. The sign of the Boltzmann
temperature plays a crucial role in this context, defining the
features of the coherent phase. Indeed, in analogy with the
single-particle-momentum distribution, it is easy to show that
\begin{equation}\label{delta_phi}
  \rho(\phi_i-\phi_{i-1})\propto \exp\left\{-\beta_B \big[1-\cos(\phi_i-\phi_{i-1})\big]\right\}.
\end{equation}
When $E\to E_M$ or $E\to 0$, the inverse temperature $\beta_B$
diverges and, depending on the sign of $\beta_B$, the distribution
Eq. \eqref{delta_phi} peaks around $\phi_i-\phi_{i-1}=0$ or $\phi_i-\phi_{i-1}=\pi$, see
Fig. \ref{fig:spatial_coherence} A.

Let us stress that not every state with negative temperature is
spatially ordered: the necessary condition is a small corresponding
phase space volume, which implies a very high energy or, equivalently,
a very small negative temperature. The same argument applies to small
positive temperatures. Of course, if negative temperatures appear,
they signal a reduction of phase space with increasing energy, and
therefore announce a more ordered structure at higher energy.

\section{Conclusions}
\label{conclusions}

In this paper we have given a series of arguments to support the
thesis of the Boltzmann temperature $T_B$ as a useful parameter to
describe the statistical features of a system with many particles and
short-range interactions, even when it takes negative values.  Let us
draw our conclusions with a series of remarks on the role of the negative
temperature and some comments on recent papers.

We have shown that the temperature $T_B$ is the proper quantity which
describes the distribution of the energy fluctuations in the canonical
ensemble.  It also enters in an immediate generalization of the
Mawell-Boltzmann distribution to the case of ``kinetic energy" which is
not a quadratic function of momentum. For a particular model we have
also demonstrated that at small $|T_B|$ (for both positive and
negative values) some kind of spatial order induced by interactions
appears, whose qualitative traits depend upon the temperature's sign.

If the microcanonical entropy $S(e)$ is a convex function, independently
of the sign of $T_B$, there is no ambiguity in determining the flux of
energy: it always goes from the hotter system, i.e. with smaller
$\beta_B$ to the colder one (with larger $\beta_B$).  It should be
reminded that the convexity of $S(e)$ can be violated only for very
small systems or systems with long range interaction, both cases being
very well known examples that can violate thermodynamic requirements.

From a physical point of view it is possible to obtain the canonical
ensemble from the microcanical one only for large systems with short
range interactions.  In such a class of systems, if $N \gg 1$, the
$S(e)$ is convex and it is easy to obtain the equivalence of the
ensembles. Such a property is a fundamental requirement to obtain
equilibrium thermodynamics, where there is no difference between
thermostatted and isolated macroscopic systems.  It is worth emphasizing that the
equivalence of the ensembles only holds if one adopts the Boltzmann
definition of entropy: for this reason, in systems exhibiting negative
temperatures, where $S_B$ and $S_G$ are no longer equivalent in the
large $N$ limit, thermodynamic can be recovered for $N\to \infty$ only
through the Boltzmann formalism.

In systems with few components and/or with long range interactions,
one can still define a canonical ensemble at a formal level
(i.e. assume that the phase space distribution is $\propto e^{-\beta
  H}$), and then wonder about the equivalence of the ensembles.
However such a formal mathematical approach, in our opinion, has no
physical meaning.  Since in presence of long range interactions (or
equivalently a system with $N=O(1)$) it is not possible to make a
clear distinction between the system and the reservoir, it is not
possible to construct systems following a canonical distribution.  For
the same reason the question of the flux of energy among two systems
appears to be meaningless in those cases.

Following Rugh~\cite{rugh1997}, $T_B$ can be computed via a molecular dynamics simulation,
and (at least in principle) from the data of an experiment.
The microcanonical formula~\eqref{eqpart}, which, in most cases, allows for a practical definition
of $T_G$, can fail in systems with negative $T_B$, therefore,  as far as we know,
at variance with $T_B$, there is not a general method to compute $T_G$ in an experiment.  

We underline that the counterexamples used in~\cite{Hilbert2014} to support
the claimed inconsistency of the use of $T_B$ are based on systems with very few degrees of freedom
and non convex $S(e)$. Let us note that the system in eq. (71)
of~\cite{Hilbert2014} is nothing but the system considered in our
Section~\ref{model}, Eq.~\eqref{modham}, with $N=1$ and $\epsilon=0$: the
claimed strange behavior of $T_B$ is present only if $N=O(1)$. On the
contrary for $N \gg 1$ as a consequence of the convexity of $S(e)$ one
has a quite natural scenario, as discussed above.  In a similar way
we have shown that the consistency of $T_G$ with the microcanonical formula fails
for large $N$.

In the microcanonical ensemble the temperature $T_B$ is a function of
the total energy $E$.  In the canonical ensemble the temperature $T_B$
is a mere property of the reservoir and does not depend on the
microscopic configuration of the system.  In~\cite{Hilbert2014},
see Sect. 3.D, the wrong concept of temperature (in non-isolated (sub)-systems) depending upon the
energy of the microscopic configuration, see their Eq.~(31), is used
to claim the inconsistency of $T_B$.  Such confusion seems to be
persistent, see~\cite{Tfluct} for a discussion of the topic of the
(non existing) fluctuations of temperature. 

In conclusion our analysis, that applies to a large class of systems
with many degrees of freedom and short-ranged interactions, shows that
the Boltzmann temperature has the following properties: i) it is the
proper quantity ruling the fluctuations of energy of a sub-system; ii)
it can be measured by means of time-averages of a suitable observable;
iii) it rules the direction of the fluxes of energies between two
coupled systems at different initial temperatures. About the Gibbs
temperature, we can mention that: i) the Gibbs entropy is an adiabatic
invariant (although a mathematically rigorous proof exists only for
one-dimensional systems); ii) the microcanonical formula for
equipartition in general is not valid therefore - at variance with
$T_B$ - a simple way to measure $T_G$ is not available. We note that
the differences between $T_B$ and $T_G$ can survive for large $N$,
even when the ensembles are equivalent in the thermodynamic limit.


\begin{acknowledgments}
  The authors acknowledge P. Buonsante, M. Cencini, M. Falcioni, U. Marini Bettolo
  Marconi and G.-L. Oppo for the many discussions and for reading the
  manuscript. We owe M. Cencini and M. Falcioni, who also contributed at a first
  stage of this work, special thanks.
\end{acknowledgments}




\bibliographystyle{apsrev4-1.bst}
\bibliography{mergedbiblio.bib}

\end{document}